\documentclass{article}

\usepackage{amsfonts}
\usepackage{amsmath}

\input{tcilatex}

\begin{document}

\author{O. Cherbal$^{(1)}$, M. Drir$^{(1)}$, M. Maamache $^{(2)}$, and D.A. Trifonov$^{(3)}$ \\[2mm]
$^{(1)}$\,{\small  Physical Faculty, Theoretical Physics Lab, USTHB,} \\
{\small B.P. 32 El-alia, Bab Ezzouar, 16111 Algiers,\ Algeria.}\\
$^{(2)}$\,{\small  Laboratoire de Physique Quantique et Syst\`{e}mes
Dynamiques, }\\
{\small Department of Physics, Setif University, Setif 19000, Algeria.}%
\\
$^{(3)}$\,{\small  Institute of nuclear research, 72 Tzarigradsko
chauss\'ee,} \\
{\small 1784 Sofia, Bulgaria.}}
\title{Fermionic coherent states for pseudo-Hermitian two-level systems}
\maketitle

\vspace{-8.05cm}
\noindent quant-ph/0608177v2
\vspace{8.05cm}

\begin{abstract}
We introduce creation and annihilation operators of pseudo-Hermitian fermions 
for two-level systems described by pseudo-Hermitian Hamiltonian with real 
eigenvalues. This allows the generalization of the fermionic coherent states 
approach to such systems. Pseudo-fermionic coherent states are constructed 
as eigenstates of two pseudo-fermion annihilation operators. These coherent
states form a bi-normal and bi-overcomplete system, and their evolution
governed by the pseudo-Hermitian Hamiltonian is temporally stable. In terms 
of the introduced pseudo-fermion operators the two-level system' Hamiltonian 
takes a factorized form similar to that of a harmonic oscillator. 
\end{abstract}

\medskip %

\section{Introduction}

In the last years a great deal of interest has been devoted to the study of
non-Hermitian\ Hamiltonians with real spectrum (see \cite{Khare}- \cite 
{Caliceti} 
and references therein). Bender and Boettcher were the first to
touch on this issue \cite{Bender1}, by introducing the notion of PT-symmetry
for one-dimensional non-Hermitian\ Hamiltonian $H_{\nu} =
p^{2}+x^{2}(ix)^{\nu }$, $(\nu \geq 0)$, that possesses real, positive and
discrete spectrum. In \cite{Dorey, Shin} the Bessis -– Zinn Justin conjecture about the
reality of the spectrum of the $PT$-symmetric Hamiltonian $-d^2/dx^2 - (ix)^{2M}$ for 
$M \geq 1$ has been proven. A criterion for the reality of the spectrum of non-Hermitian 
$PT$-symmetric Hamiltonians is provided in \cite{Caliceti}.

By definition a $PT$-symmetric Hamiltonians $H$ \ satisfies
the relation 
\begin{equation}
\left[ H, PT\ \right] = 0 \, ,  \label{1}
\end{equation}%
where $P$ and $T$\ are the operators of parity and time-reversal
transformations respectively. These are defined according to 
\begin{equation}
PxP = -x,\quad PpP = TpT = -p, \quad Ti1T = -i1 \, ,  \label{2}
\end{equation}
where $x,p,1$ are the position, momentum, and identity operators
respectively, acting on the Hilbert space, and $i:=\sqrt{-1}.$

\medskip Later, Mostafazadeh \cite{Mostafa1}-\cite{Mostafa4b} introduced the
notion of pseudo-Hermiticity in order to establish the mathematical relation
with the notion of PT symmetry. He explored the basic structure responsible
for the reality of the spectrum of non-Hermitian Hamiltonians, and
established that all the PT-symmetric non-Hermitian Hamiltonians are
pseudo-Hermitian. He also has shown that any diagonalizable operators with
discrete spectra, is pseudo-Hermitian if and only if its eigenvalues are
either real or grouped in complex-conjugate pairs (with the same
multiplicity). Moreover this result has been generalized to the class all
PT-symmetric standard Hamiltonians having $\mathbb{R}$ as their
configuration space, and to the class of possibly nondiagonalizable
Hamiltonians that admit a block-diagonalization with finite-dimensional
diagonal blocks. 
In fact  many of the later developments in the field are anticipated in the 
paper by Scholtz, Geyer and Hahne \cite{Scholtz} (see also \cite{Bender06}).

By definition {\normalsize \cite{Mostafa1}}, a Hamiltonian $H$ is called
pseudo-Hermitian if it satisfies the relation 
\begin{equation}
H^{\dagger} = \eta H\eta^{-1} \, ,  \label{3}
\end{equation}
where $\eta$ is a linear, Hermitian, and invertible operator. One can also
express the definition (\ref{3}) in the form $H^{\#} = H$, where $%
H^{\#}=\eta^{-1}H^{\dagger}\eta$ is the $\eta$-pseudo adjoint of $H$ \cite%
{Mostafa1}.

An interesting area where the pseudo-Hermiticity is applied is in the study
of non-Hermitian two-level systems \cite{Mostafa4, Aryeh}. These
simple Hamiltonian systems model accurately many physical systems in
condensed mater, atomic physics, and quantum optics \cite{Bohr} - \cite
{Choutri}. The latter field provides a beautiful implementation of the
coherent states formalism \cite{Glauber} - \cite{Twareque}. Rabi
oscillations in the non-Hermitian system of a two-level atom in
electromagnetic field have been recently examined in ref. \cite{Aryeh}.

\medskip In the preceding paper \cite{Cherbal}, we have shown how the exact
evolution and nonadiabatic Hannay's angle of Grassmannian classical
mechanics of spin one half in a varying external magnetic field, is
associates with the evolution of Grassmannian invariant-angle coherent
states.

In this paper we extend the fermionic coherent states approach 
\cite{Cahill, Lee, Abe, Ohnuki} to  two-level non-Hermitian Hamiltonians 
which are pseudo-Hermitian (\textit{p-Hermitian}). The underlying number 
system is Grassmann algebra \cite{Berezin1, Berezin2}. The set of coherent 
states (CS) for pseudo-fermionic (shortly \textit{p-fermionic}) system turned 
out to consist of two subsets of states, which are {\it bi-normalized and bi-overcomplete} 
(shortly bi-normal CS).

{\normalsize The paper is organized as follows.\thinspace\ In section 2, we
study a non-Hermitian two-level systems (a two-level atom interacting with 
electromagnetic field)and its pseudo-Hermitian properties. Then we introduce 
the creation and annihilation\ operators } for
the two-level p-Hermitian system with real energy spectrum, such that it's 
Hamiltonian ascribes a form similar to that of the free harmonic oscillator: 
$H = \Omega (b^{\#}b-1/2)$,
where $b$ and $b^{\#}$ are pseudo-fermionic (p-fermionic) lowering and
raising operators. \thinspace\ In section 3, we construct the p-fermionic CS
as eigenstates of two annihilation operators, the eigenvalues being
complex Grassmann variables. The set of such eigenstates form a bi-normal
and bi-overcomplete system. \thinspace\ Then, in section 4, we study the
time evolution of the constructed p-fermionic CS for the corresponding
two-level p-Hermitian system. This evolution is shown to be temporally stable. 
{\normalsize %
The paper ends with concluding remarks.}\ 

\medskip

\section{ Two-level systems and pseudo-Hermitian \newline
fermions}

We consider a two-level atom interacting with an electromagnetic field. The
general state of the two-level atomic system is 
\begin{equation*}
|\psi\rangle = C_{a}^\prime(t)|+\rangle + C_{b}^\prime(t)|-\rangle \, ,
\end{equation*}
where $C_{\mp}$ are the amplitudes of being in $|\mp\rangle$. They are time
dependent due to atom-field interaction. In the interaction picture (dipole
interaction and phenomenologically described decay), and in the rotating
wave approximation, the evolution of the system is described by the equation 
\cite{Lamb, Garrison}

\begin{equation}
i\frac{\partial }{\partial t}\left( 
\begin{array}{c}
C_{a}^\prime(t) \\ 
C_{b}^\prime(t)%
\end{array}
\right) =\frac{1}{2}\left( 
\begin{array}{cc}
-i\gamma _{a} & \omega ^{\ast } \\ 
\omega & -i\gamma _{b}%
\end{array}
\right) \left( 
\begin{array}{c}
C_{a}^\prime(t) \\ 
C_{b}^\prime(t)%
\end{array}
\right) \, .  \label{4}
\end{equation}
The real constants $\gamma _{a}$, $\gamma _{b}$ are the decay rates for the
upper and lower level respectively. The quantity $\omega $ characterizes the
radiation-atom interaction matrix element between the levels ( $\omega
^{\ast }$ is the complex conjugate ). The basic vectors of the upper (lower
) level are $\mid +\rangle$ \ and $\mid -\rangle$ .

We remove the average effect of the decay terms by means of a nonunitary
transformation in the state space,

\begin{equation}
|\psi\rangle \rightarrow U(t) |\psi\rangle, \quad U(t) = e^{\Gamma t},\,\,
\Gamma = \frac 14(\gamma_a + \gamma_b).  \label{5}
\end{equation}

\noindent The probability amplitudes in the new representation are $C_i(t) =
\exp(\Gamma t)C_i^\prime(t)$, $i=a,b$, and the non-Hermitian Hamiltonian
takes the following matrix form

\begin{equation}
H = \frac{1}{2}\left( 
\begin{array}{cc}
-i\delta & \omega ^{\ast } \\ 
\omega & i\delta%
\end{array}%
\right) \, ,  \label{6}
\end{equation}%
where $\delta = (\gamma _{a}-\gamma _{b})/2$.

The trace of $H$, eq.(\ref{4}), is vanishing, and the determinant of $H$ is
real, $\det H=(\delta ^{2}-|\omega |^{2})/4$. Therefore it is $\eta $%
-pseudo-Hermitian ($\eta $-p-Hermitian) \cite{Mostafa3}. Its matrix is a
particular case of a more general $2\times 2$ traceless matrix studied in 
\cite{Mostafa4}, where the complete biorthonormal system $\{|\psi
_{i}\rangle ,|\phi _{i}\rangle \}$ for $H$ and the operator $\eta $ are
explicitly constructed. We reproduce this system and $\eta $ (up to certain
common factors) in our specific notations.

The eigenvalues $E_{i}$ of $H$, $i=1,2$, and the related complete
biorthonormal system are given by (we consider the nondegenerate case of $%
E_i\neq 0$)

\begin{equation}
E_{1} = -\frac{\Omega}{2}\,\, ,\qquad E_{2}=\frac{\Omega}{2}\,\, ,  \label{7}
\end{equation}

\begin{equation}
\left\vert \psi_{1}\right\rangle = \,\,\frac{1}{\sqrt{2\Omega}}\,\left( 
\begin{array}{c}
\frac{-\omega^{\ast}\sqrt{\Omega+i\delta}}{\left\vert \omega\right\vert } \\ 
\\ 
\sqrt{\Omega-i\delta}%
\end{array}
\right) ,\text{ \ \ \ \ }\left\vert \psi_{2}\right\rangle = \, \frac{1}{%
\sqrt{2\Omega}} \, \left( 
\begin{array}{c}
\frac{\omega^{\ast}\sqrt{\Omega-i\delta}}{\left\vert \omega\right\vert } \\ 
\\ 
\sqrt{\Omega+i\delta}%
\end{array}
\right) \, ,  \label{8}
\end{equation}


\begin{equation}
\left\vert \phi _{1}\right\rangle =\text{\ \ }\frac{1}{\sqrt{2\Omega ^{\ast }%
}}\text{\ }\left( 
\begin{array}{c}
\frac{-\omega ^{\ast }\sqrt{\Omega ^{\ast }-i\delta }}{\left\vert \omega
\right\vert } \\ 
\\ 
\sqrt{\Omega ^{\ast }+i\delta }%
\end{array}%
\right) ,\text{ \ \ \ \ }\left\vert \phi _{2}\right\rangle =\,\frac{1}{\sqrt{%
2\Omega ^{\ast }}}\,\left( 
\begin{array}{c}
\frac{\omega ^{\ast }\sqrt{\Omega ^{\ast }+i\delta }}{\left\vert \omega
\right\vert } \\ 
\\ 
\sqrt{\Omega ^{\ast }-i\delta }%
\end{array}%
\right) ,  \label{9}
\end{equation}%
\medskip where $\ \Omega =\sqrt{\left\vert \omega \right\vert ^{2}-\delta
^{2}}.$

For both real and complex eigenvalues (i.e. real and complex $\Omega$) the Hamiltonian 
(\ref{6}) satisfies the p-Hermiticity relation (\ref{3}) with $\eta $ given by

\begin{equation}
\eta = \left( 
\begin{array}{cc}
1 & \frac{i\delta\omega^{\ast}}{\left\vert \omega\right\vert ^{2}} \\ 
-\frac{i\delta\omega}{\left\vert \omega\right\vert ^{2}} & 1%
\end{array}
\right) \, .  \label{10}
\end{equation}

As noted in \cite{Aryeh}, the real eigenvalues correspond to the case where
the dipole interaction is large relative to the damping effect. In this case
the ordinary Rabi frequency is replaced by the 'pseudo Rabi frequency' which
in our notations are $|\omega |/2$ and $\Omega /2$ correspondingly. 


In the case of $\Omega = 0$ (i.e. $|\omega|^2 = \delta^2$), the amplitudes 
$C_a^\prime$,$C_b^\prime$ are given \cite{Aryeh} by 
$(1 -\delta t) \exp(-\Gamma t)$ and $(i\omega t/2) \exp(-\Gamma t)$ correspondingly, 
where $ \Gamma = (\gamma_a + \gamma_b)/4$. Due
to the exponential decay factor $\exp(-\Gamma t)$, any divergence [27] does not occur 
in our system.

In the case of $|\omega|^2 < \delta^2$ the eigenvalues of $H$ are pure imaginary, 
but the Hamiltonian is still pseudo-Hermitian \cite{Aryeh}.
The $PT$-symmetry of our Hamiltonian (6) is considered in \cite{Aryeh} following the method of
Bender, Brody and Jones \cite{Bender2}. The Parity operator $P$ of the two-level system can be defined as
\cite{Bender2}
\[ P = \left(\begin{matrix}0 & 1\cr 1& 0 \end{matrix}\right),\] 
so that $P^2 = 1$, $P = P^{-1}$.

The generalized parity operator for the two-level systems is defined \cite{Mostafa4b, Ahmed} as
$P = |\phi_1\rangle\langle\phi_1| - |\phi_2\rangle\langle\phi_2|$, which in our case results to
\[P = \left( 
\begin{matrix}0 & -\frac{\omega^*}{|\omega|}\cr -\frac{\omega}{|\omega|}& 0 \end{matrix} 
\right) .\]

Different definitions have been introduced by Mostafazadeh \cite{Mostafa4b} and Ahmed 
\cite{Ahmed} for the antilinear time-reversal operator $T$. As in the paper \cite{Aryeh}, 
we use the representation introduced by Bender et al \cite{Bender2}, namely,
\[ T = K_0, \]
where $K_0$ is the complex conjugation operator. One has $K_0^2 = 1$, $(PT)^2 = 1$, and one finds
that $PT$ commutes with $H$:\, $PK_0HK_0^{-1}P^{-1} = H$. If one uses the generalized parity operator
$P = |\phi_1\rangle\langle\phi_1| - |\phi_2\rangle\langle\phi_2|$, then one can take 
$T = UK_0$, where $U$ is a $2\times 2$ unitary diagonal
matrix, $U_{22} = U_{11} = \omega/|\omega|$.

The generalized charge conjugation operator $C$ for two-level system is given by the
expression \cite{Mostafa4b} $C = |\psi_1\rangle\langle\phi_1| - |\psi_2\rangle\langle\phi_2|$ ,
 which in our case reads

\[ C = \frac{1}{2\Omega}\left(\begin{matrix} i\delta & -\omega^*\cr -\omega &-i\delta 
\end{matrix} \right)  = -\frac{2}{\Omega} H.\]

From the last equation we deduce that $C$ commutes with the Hamiltonian $H$:\,  $[C,H] = 0$.
This invariance property eliminates negative inner products \cite{Aryeh}. 


Furthermore, unless otherwise stated, we consider the small damping effect case of our
system, i.e. $\omega^2 > \delta^2$, that is $\Omega $ real (and if real it 
is positive). One can verify that, in this regime (i.e. $|\omega |^{2}>\delta ^{2}$) 
this $\eta $, eq. (10), can be represented in terms of $|\phi _{i}\rangle $ as 

\begin{equation}  \label{11}
\eta = \frac{\Omega}{|\omega|} \left(|\phi_1\rangle\langle \phi_1| +
|\phi_2\rangle\langle \phi_2|\right) \equiv \frac{\Omega}{|\omega|}\eta_+ \, .
\end{equation}

The operator $\eta$ is positive definite. The construction $\eta_+ =
|\phi_1\rangle\langle \phi_1| + |\phi_2\rangle\langle \phi_2|$ was
introduced by Mostafazadeh \cite{Mostafa5}.

Now, let us introduce the annihilation operator $b$ associated to the
Hamiltonian $H$ given in eq. (\ref{6}),

\begin{equation}
b = \frac{1}{2\Omega}\left( 
\begin{array}{cc}
-\left\vert \omega\right\vert & \frac{-\omega^{\ast}(\Omega+i\delta )}{%
\left\vert \omega\right\vert } \\ 
\frac{\omega(\Omega-i\delta)}{\left\vert \omega\right\vert } & \left\vert
\omega\right\vert%
\end{array}
\right) \, .  \label{12}
\end{equation}
Its adjoint operator reads ($\Omega$ is real)

\begin{equation}
b^{+}=\frac{1}{2\Omega }\left( 
\begin{array}{cc}
-\left\vert \omega \right\vert & \frac{\omega ^{\ast }(\Omega +i\delta )}{%
\left\vert \omega \right\vert } \\ 
\frac{-\omega (\Omega -i\delta )}{\left\vert \omega \right\vert } & 
\left\vert \omega \right\vert%
\end{array}%
\right) \,,  \label{13}
\end{equation}%
and its $\eta $-p-Hermitian adjoint $b^{\#}$, defined by

\begin{equation}
b^{\#} = \eta ^{-1}b^{+}\eta \, ,  \label{14}
\end{equation}%
takes the form

\begin{equation}
b^{\#} = \frac{1}{2\Omega}\left( 
\begin{array}{cc}
-\left\vert \omega\right\vert & \frac{\omega^{\ast}(\Omega-i\delta )}{%
\left\vert \omega\right\vert } \\ 
\frac{-\omega(\Omega+i\delta)}{\left\vert \omega\right\vert } & \left\vert
\omega\right\vert%
\end{array}
\right) \, .  \label{15}
\end{equation}

Next we examine the properties of the operators $b^{\#}$ and $b$. First of
all is that $b^{\#}$ and $b$ realize a pseudo-Hermitian generalization of
the fermion algebra, namely 
\begin{equation}
b^{2}=b^{\#2}=0,\text{ \ \ }\left\{ b,b^{\#}\right\} =bb^{\#}+b^{\#}b=1\text{%
\ \ }.  \label{16}
\end{equation}%
$b^{\#}$ and $b$ could be called the creation and annihilation\ operators of
p-Hermitian fermions \cite{Mostafa5}. One can verify that they raise and
lower the eigenvalues of $H$ by a quantity $\Omega =2E$, i.e. they act on
the states $|\psi _{i}\rangle $ as follows

\begin{equation}
b\left\vert \psi_{1}\right\rangle = 0,\text{\ \ \ \ \ \ \ \ }b\left\vert
\psi_{2}\right\rangle =\left\vert \psi_{1}\right\rangle \, ,  \label{17}
\end{equation}

\begin{equation}
b^{\#}\left\vert \psi_{2}\right\rangle = 0,\qquad b^{\#}\vert \psi_1\rangle
= \vert \psi_2\rangle \, ,  \label{18}
\end{equation}%
The operator $b$ annihilates the lowest eigenstates $\vert\psi_{1}\rangle$,
and $b^{\#}$ brings this state onto the upper eigenstates $\vert
\psi_{2}\rangle$.

Introducing the quadratic operator $N=b^{\#}b$, the p-fermionic number
operator, we find the following natural anticommutation relations

\begin{equation}
\left\{N, b\right\} = b \qquad \left\{ N,b^{\#}\right\} = b^{\#} \text{ \ \ }
.  \label{19}
\end{equation}
In terms of the operators $b$ and $b^{\#}$ the Hamiltonian $H$\ is
factorized (up to an additive $C$-number term) to a form, similar to that of
the free (boson) harmonic oscillator,

\begin{equation}
\ H = \Omega\left( b^{\#}b - \frac{1}{2}\right) \, .  \label{20}
\end{equation}

Taking the Hermitian conjugate of both sides of (\ref{20}) we reaffirm the
p-Hermiticity of $H$ (according to definition (\ref{3})):

\begin{align}
H^{+} & =\Omega(b^{+}\eta b\eta^{-1}-\frac{1}{2})  \notag \\
& =\Omega\eta\eta^{-1}(b^{+}\eta b\eta^{-1}-\frac{1}{2})\eta\eta ^{-1} 
\notag \\
& =\eta H\eta^{-1} \, .  \notag
\end{align}

The above relations confirm that $\ b\ $ and $\ b^{\#}$ \ are lowering and
raising operators for the two-level p-Hermitian system (with real
eigenvalues) and can be regarded as p-fermionic annihilation and creation
operators. This is consistent with the limit $\delta =0$, corresponding to a
Hermitian Hamiltonian, when $\eta =1$ and $b^{\#}=b^{+}$, i.e.\ the
p-Hermitian generalization of the fermion algebra reduces to the usual
fermion algebra. Quantum system with Hamiltonian of the form (\ref{20})
should be referred to as {\it p-fermionic oscillator}.

\medskip

\section{ Pseudo-fermionic coherent states}

\medskip Having introduced the p-fermion lowering and raising operators we
now embark on the construction of the p-fermionic coherent states (CS) for
our system described by p-Hermitian Hamiltonian $H$ given in eq. (\ref{6}).
We shall follow as closely as possible the scheme of fermionic CS developed 
in papers \cite{Cahill, Lee, Abe, Ohnuki}, generalizing it to the p-fermion 
case.

\noindent For the reader convenience we begin with a brief reminder of the
properties of complex Grassmann variables \cite{Cahill, Abe, Maam, Ohnuki},
denoted here as $\xi $ and $\xi ^{\ast }$.

The complex Grassmannian variables $\xi_{i}$ and their complex conjugates $%
\xi_{i}^{\ast}$ satisfy the anticommutation relations:

\begin{equation}
\left\{ \xi_{i},\xi_{j}\right\} =\xi_{i}\xi_{j}+\xi_{j}\xi_{i} = 0\,,
\label{21}
\end{equation}%
\begin{equation}
\left\{ \xi_{i}^{\ast},\xi_{j}\right\} =0,\text{ \ \ }\left\{ \xi_{i}^{\ast},%
\text{ }\xi_{j}^{\ast}\right\} =0\text{\ }.  \label{22}
\end{equation}
The $\xi_{i}$ 's anticommute with $b$ \ and \ $b^{\#}$,

\begin{equation}
\left\{ \xi_{i},b\right\} =0,\text{ }\left\{ \xi_{i}^{\ast},b\right\} = 0\,, 
\text{\ \ }\left\{ \xi_{i},b^{\#}\right\} =0\text{\ } ,  \label{23}
\end{equation}
and have the following properties%
\begin{align}
\xi |\psi_1\rangle &= |\psi_1\rangle \xi\,,\qquad \xi|\psi_2\rangle =
-|\psi_2\rangle \, ,  \label{24} \\
\xi |\phi_1\rangle &= |\phi_1\rangle \xi\,, \qquad \xi|\phi_2\rangle =
-|\phi_2\rangle \xi \text{\ } .  \label{25}
\end{align}
The pseudo-Hermitian conjugation reverses the order of all fermionic
quantities, both the operators and the Grassmann variables:

\begin{equation}
(b^{\#}\xi _{i}+\xi _{i}^{\ast }b)^{\#}=\xi _{i}^{\ast }b+b^{\#}\xi _{i} \,.
\label{26}
\end{equation}%
The Grassmann integration and differentiation over the complex Grassmann
variables are given by

\begin{equation}
\int d\xi\text{ }1=0\text{, \ \ }\int d\xi\xi=1,\text{ }\int d\xi^{\ast}%
\text{ }1=0\text{, \ \ }\int d\xi^{\ast}\xi^{\ast} = 1 \,,  \label{27}
\end{equation}

\begin{equation}
\frac{d}{d\xi}1 = 0\text{, \ \ }\frac{d}{d\xi}\xi=1,\text{ \ }\frac{d}{%
d\xi^{\ast}}1 = 0\text{, \ \ }\frac{d}{d\xi^{\ast}}\xi^{\ast} = 1\,.
\label{28}
\end{equation}
The Grassmann integration of any function is equivalent to the left
differentiation

\begin{equation}
\int d\xi\text{ }f(\xi)=\frac{\partial}{\partial\xi}\text{ }f(\xi) \, .
\label{29}
\end{equation}

We define the displacement operators $D(\xi)$ for any set of complex
Grassmannian variables $\xi$ in the following way:

\begin{align}
D(\xi )& =\exp \left( b^{\#}\xi -\xi ^{\ast }b\right)  \label{30} \\
& = 1+b^{\#}\xi -\xi ^{\ast }b+\left( b^{\#}b-\frac{1}{2}\right) \xi ^{\ast
}\xi \, .  \label{31}
\end{align}

\medskip The pseudo-Hermitian adjoint $D^{\#}$ is given by

\begin{align}
D^{\#}(\xi )& =\exp \left( \xi ^{\ast }b-b^{\#}\xi \right)  \label{32} \\
& = 1+\xi ^{\ast }b-b^{\#}\xi +\left( b^{\#}b-\frac{1}{2}\right) \xi ^{\ast
}\xi \, .  \label{33}
\end{align}
These two operators satisfy the following displacement relation, 
\begin{equation}
D^{\#}(\xi ) b D(\xi ) = b + \xi\, .  \notag
\end{equation}

\medskip Using the explicate formulas of $D$ and $D^\#$, and the
anticommutation relations between operators $b$, $b^\#$ and Grassmann
variable $\xi$ we establish that $D(\xi)$ are pseudo-unitary: $%
D^{\#}(\xi)D(\xi) = 1 = D(\xi)D^{\#}(\xi)$.

We now define the p-fermionic CS $\left\vert \xi \right\rangle $ as
eigenstates of the annihilation operator $b$,

\begin{equation}
b\left\vert \xi\right\rangle =\xi\left\vert \xi\right\rangle \, .  \label{34}
\end{equation}
The eigenvalue $\xi$ is a complex Grassmannian variable.

The Hermitian adjoint of the CS (the bra-vector) is $\langle \xi\vert $ and
it is left eigenstate of $b^\dagger$, $\langle \xi\vert b^{\dagger} =
\langle \xi\vert \xi^{\ast}$. In order to meet the alternative relation $%
_\eta\langle \xi\vert b^{\#} = \,_\eta\langle \xi\vert \xi^{\ast}$ one has
to define $_\eta\langle \xi\vert \equiv (|\xi\rangle)^{\#} := \langle
\xi|\eta$.

Similarly to the cases of Glauber bosonic CS \cite{Glauber} and of fermionic
CS \cite{Cahill} our p-fermion eigenstates $|\xi \rangle $ can be
constructed from the lowest (ground) eigenstate $|\psi _{1}\rangle $ of the
Hamiltonian $H$, acting on it by the pseudo-unitary operator $D(\xi )$:

\begin{equation}
\left\vert \xi \right\rangle = D(\xi)\left\vert \psi_{1}\right\rangle\, .
\label{35}
\end{equation}%
By using the formula (\ref{31}) for the displacement operator, we may write
the state $\left\vert \xi \right\rangle $ in the form

\begin{equation}
\left\vert \xi\right\rangle =\exp\left( -\frac{1}{2}\xi^{\ast}\xi\right)
\left( \left\vert \psi_{1}\right\rangle -\text{ }\xi\left\vert \psi
_{2}\right\rangle \right) \, .  \label{36}
\end{equation}
The Hermitian adjoint of the CS is

\begin{align}
\left\langle \xi\right\vert & =\left\langle \psi_{1}\right\vert \mid
D^{\dagger}(\xi)  \label{37} \\
& = \exp\left( -\frac{1}{2}\xi^{\ast}\xi\right) \left(\langle \psi_{1}\mid 
+ \xi^{\ast}\langle \psi_{2}\vert \right)\, ,  \label{38}
\end{align}
and the inner product $\langle\xi|\xi\rangle$ is

\begin{equation}  \label{39}
\langle\xi|\xi\rangle = \langle \psi_1|\psi_1\rangle + (\langle
\psi_2|\psi_2\rangle - \langle \psi_1|\psi_1\rangle)\xi^*\xi - 2i\mathrm{Im}
(\xi\langle\psi_1|\psi_2\rangle)\, ,
\end{equation}
so that the $\vert \xi\rangle $ are not normalized.

Now we have to examine for (over)completeness the set $\{|\xi \rangle \}$.
One can straightforwardly check (using the rules (\ref{21}) - (\ref{27}))
that neither the integral (against the measure $d\xi ^{\ast }d\xi $) of the
Hermitian $|\xi \rangle \langle \xi |$ nor the integral of p-Hermitian $|\xi
\rangle \,_{\eta }\langle \xi |$ (unnormalized) projectors result in the
identity operator:

\begin{equation}  \label{40}
\int d\xi^* d\xi |\xi\rangle\langle \xi| \neq 1,\quad \int d\xi^* d\xi
|\xi\rangle\,_\eta\langle \xi| \neq 1 \, .
\end{equation}

The way out of this impasse is suggesting by the known transition from
'orthonormal system' of eigenstates of Hermitian $H$ to the 'biorthonormal
system' of states of p-Hermitian $H$. With this idea in mind we introduce
another continuous family of states namely the eigenstates $\widetilde{|\xi
\rangle }$ of the operator $\tilde{b}$, that annihilates the dual state $%
|\phi _{1}\rangle $,

\begin{equation}
\tilde{b}\widetilde{|\xi \rangle }=\xi \widetilde{|\xi \rangle },\quad 
\tilde{b}=\eta b\eta ^{-1}\,.  \label{41}
\end{equation}%
Operator $\tilde{b}$ is nilpotent, $\tilde{b}^{2}=0$ and anticommutes with $%
b^{\dagger }$. Representing $b^{\dagger }=\eta \tilde{b}^{\dagger }\eta
^{-1} $ we see that $b^{\dagger }$ is $\eta ^{\prime }$-p-Hermitian adjoint
to $\tilde{b}$, $\eta ^{\prime }=\eta ^{-1}$. Denoting this
pseudo-conjugation by $^{\#\prime }$ we obtain the pair of p-fermionic
operators $\tilde{b}$ and $\tilde{b}^{\#\prime }$,

\begin{equation}  \label{42}
\tilde{b}\,\tilde{b}^{\#\prime} + \tilde{b}^{\#\prime}\tilde{b} = 1, \quad 
\tilde{b}^2 = (\tilde{b}^{\#\prime})^2 = 0\,.
\end{equation}

In view of the p-fermionic algebra (\ref{42}) we introduce new displacement
operators 
\begin{equation*}
\widetilde{D}(\xi )=\exp (\tilde{b}^{\#\prime }\xi -\xi ^{\ast }\tilde{b}%
),\quad \widetilde{D}^{\#\prime }(\xi )\,\tilde{b}\,\widetilde{D}(\xi )=%
\tilde{b}+\xi \,,
\end{equation*}%
and construct eigenstates of $\tilde{b}$ according to the above described
scheme (see eqs. (\ref{35}), (\ref{36})),

\begin{align}
\widetilde{\vert \xi\rangle} &= \widetilde{D}(\xi) \vert\phi_{1}\rangle
\label{43} \\
&=\exp\left( -\frac{1}{2}\xi^{\ast}\xi\right) \left(\vert \phi_{1}\rangle -
\xi\vert \phi_{2}\rangle \right)\, .  \label{44}
\end{align}

The scalar product between $\widetilde{\langle\xi|}\widetilde{\xi\rangle}$
takes the form

\begin{equation}  \label{45}
\widetilde{\langle\xi|}\widetilde{\xi\rangle} = \langle \phi_1|\phi_1\rangle
+ (\langle \phi_2|\phi_2\rangle - \langle \phi_1|\phi_1\rangle)\xi^*\xi - 2i%
\mathrm{Im} (\xi\langle\phi_1|\phi_2\rangle),
\end{equation}
while 
\begin{equation*}
\widetilde{\langle\xi|}\xi\rangle = \frac{|\omega|}{\Omega}%
\langle\xi|\eta|\xi\rangle = 1,
\end{equation*}
or more generally,

\begin{equation}  \label{46}
\langle\xi|\widetilde{\zeta\rangle} = \langle \psi_1\vert D^\dagger(\xi)%
\widetilde{D}(\zeta)\vert\phi_1\rangle = \xi^*\zeta + \frac
14(2-\xi^*\xi)(2-\zeta^*\zeta).
\end{equation}

By means of the two type of states $|\xi\rangle$ and $\widetilde{|\xi\rangle}
$ the resolution of the identity is realized in the following way, 
\begin{equation}  \label{47}
1 = \int d\xi^*d\xi \,\vert\xi\rangle\widetilde{\langle\xi\vert} = \int
d\xi^*d\xi \,\widetilde{\vert\xi\rangle}\langle\xi\vert.
\end{equation}
The equations (\ref{47}) can be easily verified using the expansions of $%
|\xi\rangle$ and $\widetilde{|\xi\rangle}$ in terms of $|\psi_i\rangle$ and $%
|\phi_i\rangle$ (eqs. (\ref{36}) and (\ref{44})) and the rules of
permutation and integration (\ref{21}) - (\ref{27}).

We have obtained that the system of one-mode \textit{p-fermionic CS} consists
of two subsets, namely $\{|\xi \rangle \}$ and $\{|\widetilde{\xi \rangle }%
\}$. In view of (\ref{45}) and (\ref{46}) this continuous system should be
called \textit{bi-normalized and bi-overcomplete}, or shortly system of 
\textit{bi-normal CS}. Similarly the two sets of pseudo-unitary operators $%
D(\xi )$, $\widetilde{D}(\xi )$ should be called \textit{bi-unitary}: 
\begin{equation*}
D(\xi )\widetilde{D}^{\dagger }(\xi )=1=\widetilde{D}^{\dagger }(\xi )D(\xi
).
\end{equation*}%
Note that $D(\xi )$ is $\eta $-pseudo-unitary, while $\widetilde{D}(\xi )$
is $\eta ^{\prime }$-pseudo-unitary with $\eta ^{\prime }=\eta ^{-1}$.

\medskip

\section{Time evolution of p-fermionic coherent states}

A given parametric set of states is said to be realizable for a physical
system if the time evolution $|\psi;t\rangle$ of any initial state $%
|\psi\rangle$ from the set, governed by the Hamiltonian, leaves the state in
the set \cite{Trif}. In other word $|\psi;t\rangle$, for any $t$, obeys the
defining criteria of the set. In such a case one shortly says that the time
evolution (of the parametric set of states) is \textit{stable} \cite{Trif}.
In Hermitian mechanics this means that the time dependence of the states is
included, up to a phase factor, in the state parameters. For example the
time evolution $|\alpha;t\rangle$ of Glauber CS $|\alpha\rangle$ \cite%
{Glauber} is stable with respect to the harmonic oscillator evolution
operator $\exp(-iHt)$, $H=\omega(a^\dagger a + 1/2)$:

\begin{equation}  \label{48}
|\alpha;t\rangle = e^{-i\omega t/2}|\alpha(t)\rangle,\quad a|\alpha;t\rangle
= \alpha(t)|\alpha(t)\rangle, \quad \alpha(t) = \alpha e^{-i\omega t} \, .
\end{equation}

In the case of our p-fermionic CS $\{|\xi \rangle ,\,\widetilde{|}\xi
\rangle \}$ the set' parameter is the complex Grassmann variable $\xi $, the
eigenvalue of the p-fermionic lowering operators $b$ or $\tilde{b}$. The
time evolution is stable if the evolved states $|\xi;t\rangle $ and $%
\widetilde{|\xi;t\rangle}$ remain eigenstates of the operators $b$ and $%
\tilde{b}$ respectively,

\begin{equation}  \label{49}
b|\xi;t\rangle = \xi(t)|\xi;t\rangle, \qquad \tilde{b}\widetilde{%
|\xi;t\rangle} = \xi(t)\widetilde{|\xi;t\rangle}\, .
\end{equation}
This implies that the time evolved CS $\{|\xi;t\rangle$ and 
$\widetilde|\xi;t\rangle\}$ should form bi-normal and bi-overcomlete system.

Let us first consider the time evolution of an initial CS $|\xi\rangle$.
Clearly we have $|\xi;t\rangle = \exp(-iHt)|\xi\rangle$, $|\xi;0\rangle
\equiv |\xi\rangle$. Using the form (\ref{36}) of $|\xi\rangle$ and the
facts that $|\psi_{1,2}\rangle$ are eigenstates of $H$ (with eigenvalues $%
E_{1,2}$) we get

\begin{equation}  \label{50}
\vert \xi;t\rangle = e^{-iE_1t}\left(1 - \frac
12\xi^*\xi\right)|\psi_1\rangle - e^{-iE_2t}\xi |\psi_2\rangle \, .
\end{equation}
Taking into account that $E_1 = -E$ and $E_2 = E$ we put $\xi(t) =
e^{-i2Et}\xi$ and rewrite the last equation in the form

\begin{equation}  \label{51}
\vert \xi;t\rangle = e^{iEt}\left[\left(1 - \frac 12\xi(t)^*\xi(t)\right)
|\psi_1\rangle - \xi(t) |\psi_2\rangle\right] = e^{iEt}\vert \xi(t)\rangle
\, .
\end{equation}
which manifests the stability of the time evolution of CS $|\xi\rangle$.
Note that the overall time dependent factor $\exp(iEt)$ is a phase factor
since $E_i$ are real.

In a similar manner we establish, that the time evolution $\widetilde{%
|\xi;t\rangle}$ of an initial $\widetilde{|\xi\rangle}$, is stable (remains
eigenstate of $\tilde{b}$):

\begin{align}
\widetilde{|\xi ;t\rangle }& =\exp (-iH^{\dagger }t)\widetilde{|\xi \rangle }
\notag \\
& =\exp (-iE_{1}t)\left( 1-\frac{1}{2}\xi ^{\ast }\xi \right) |\phi
_{1}\rangle - \exp (-iE_{2}t)\xi |\phi _{2}\rangle  \notag \\
& =\exp (iEt)\left( \left( 1-\frac{1}{2}\xi (t)^{\ast }\xi (t)\right) |\phi
_{1}\rangle - \xi (t)|\phi _{2}\rangle \right) =\exp (iEt)|\xi (t)\rangle \,.
\label{52}
\end{align}%
%
The results (\ref{51}) and (\ref{52}) reveal the bi-normality and
bi-overcompleteness of the family of time evolved states $\{|\xi ;t\rangle ,\,%
\widetilde{|\xi ;t\rangle }\}$ of the p-fermionic oscillator system (\ref{20}%
): one has $\langle t;\xi \widetilde{|\xi ;t\rangle }=1$, and 
\begin{equation}
1=\int d\xi ^{\ast }d\xi |\xi ;t\rangle \widetilde{\langle t;\xi |}=\int
d\xi ^{\ast }d\xi \widetilde{|\xi ;t\rangle }\langle t;\xi |\,.  \label{53}
\end{equation}%
We observe that here the time evolved states $|\xi ;t\rangle $ and $%
\widetilde{|\xi ;t\rangle }$ differ from CS $|\xi (t)\rangle $ and $%
\widetilde{|\xi (t)\rangle }$ in phase factors only. In more general cases
the overall factors $\mathcal{N}(t)$ and $\widetilde{\mathcal{N}}(t)$ ascribed
in the stable evolution of bi-normal and bi-overcomplete system of states
could not be phase factors, but their product should equal unity, $\mathcal{N%
}^{\ast }(t)\widetilde{\mathcal{N}}(t)=1$.

Finally we have to note that a complementary bi-normal and bi-overcomplete
system of states can be constructed, in a symmetrical manner using the
operators $b^{\#}$ and $\widetilde{b^{\#}}$, that annihilate the
'upper-level states' $|\psi_2\rangle$ and $|\phi_2\rangle$.

\medskip

\section{\ Concluding Remarks}

\ \ \ \ \ In this paper, we have generalized the fermionic coherent states
(CS) for two-level systems described by pseudo-Hermitian Hamiltonian with
real spectrum. Unlike the standard bosonic and fermionic cases the system of
pseudo-fermionic (p-fermionic) CS consists of two subsets, which are
bi-normalized and bi-overcomplete. In this sense the system of p-fermionic
CS can be regarded as a continuous analogue of the bi-orthonormal system of
discrete eigenstates of p-Hermitian $H$. The two subsets are built up as
eigenstates of the p-fermion annihilation operators $b$ and $\tilde{b}=\eta
b\eta ^{-1}$, where $\eta $ is the Hermitian operator that ensures the
p-Hermiticity of the Hamiltonian, $H=\eta ^{-1}H^{\dagger }\eta $. In terms
of $b$ and $b^{\#}=\eta ^{-1}b^{\dagger }\eta $ the Hamiltonian is
factorized to the form of p-fermionic oscillator, eq. (\ref{20}).

The evolution of the p-fermionic CS governed by the p-Hermitian two-level
Hamiltonian (\ref{6}) is shown to be time stable - the evolved states remain
eigenstates of the p-fermionic annihilation operators, preserving the
bi-normality and bi-overcompleteness of the system at later time. In the
Hermitian limit of $\eta =1$ (that is $\delta =0$ in (\ref{6}) our
p-fermionic CS and all related formulas recover standard fermionic CS of
refs. \cite{Cahill, Lee}. Time evolution of fermionic CS for a Pauli spin 
in a slowly varying magnetic field was examined by Abe \cite{Abe} (see also 
the comment \cite{Maam}).

\subsection*{Acknowledgments}

One of the authors (O.C.) \ acknowledges Drs M. Le Bellac and J.P. Provost
for their useful comments and valuable discussions.

\end{document}